\documentclass[12pt,a4paper]{article}
\usepackage{amsmath}
\usepackage{amssymb}
\usepackage{cite}

\renewcommand{\thesection}
  {\arabic{section}.\hspace{-.5em}}
\renewcommand{\thesubsection}
  {\arabic{section}.\arabic{subsection}.\hspace{-.5em}}
\renewcommand{\thesubsubsection}
  {\arabic{section}.\arabic{subsection}.\arabic{subsubsection}
   \hspace{-.5em}}

\makeatletter
\renewcommand\section{
  \@startsection{section}{3}{\z@}%
  {-3.25ex\@plus -1ex \@minus -.2ex}%
  {1.5ex \@plus .2ex}%
  {\normalfont\normalsize\bfseries\mathversion{bold}}}
\renewcommand\subsection{
  \@startsection{subsection}{3}{\z@}%
  {-3.25ex\@plus -1ex \@minus -.2ex}%
  {1.5ex \@plus .2ex}%
  {\normalfont\normalsize\bfseries\mathversion{bold}}}
\renewcommand\subsubsection{
  \@startsection{subsubsection}{3}{\z@}%
  {-3.25ex\@plus -1ex \@minus -.2ex}%
  {1.5ex \@plus .2ex}%
  {\normalfont\normalsize\bfseries\mathversion{bold}}}
\makeatother

\makeatletter \@addtoreset{equation}{section} \makeatother
\renewcommand{\theequation}{\arabic{section}.\arabic{equation}}

\makeatletter
\renewcommand{\appendix}{
\renewcommand{\thesection}{\Alph{section}.\hspace{-.5em}}
\renewcommand{\thesubsection}
  {\Alph{section}.\arabic{subsection}.\hspace{-.5em}}
\renewcommand{\thesubsubsection}
  {\Alph{section}.\arabic{subsection}.\arabic{subsubsection}.
   \hspace{-.5em}}
\@addtoreset{equation}{section}
\renewcommand{\theequation}{\Alph{section}.\arabic{equation}}
\setcounter{section}{0}}
\makeatother

\newcommand{\Eqn}[1]{&\hspace{-0.5em}#1\hspace{-0.5em}&}
\newcommand{\nn}{\nonumber}
\renewcommand{\[}{\begin{equation}}
\renewcommand{\]}{\end{equation}}
\newcommand{\eqb}{\begin{eqnarray}}
\newcommand{\eqe}{\end{eqnarray}}

\newcommand{\grp}[1]{\mathrm{#1}}
\newcommand{\bvec}[1]{\boldsymbol{#1}}

\newcommand{\bbC}{{\mathbb C}}
\newcommand{\bbR}{{\mathbb R}}
\newcommand{\bbZ}{{\mathbb Z}}

\newcommand{\varth}{\vartheta}
\newcommand{\iPsi}{\mathnormal{\Psi}}

\newcommand{\diag}{\operatorname{diag}}
\newcommand{\sn}{\operatorname{sn}}
\newcommand{\cn}{\operatorname{cn}}
\newcommand{\dn}{\operatorname{dn}}

\newcommand{\pres}{\vphantom{I}\hspace{-1ex}}

\begin{document}


\def\papertitlepage{\baselineskip 3.5ex \thispagestyle{empty}}
\def\preprinumber#1#2{\hfill
\begin{minipage}{1.0in}
#1 \par\noindent #2
\end{minipage}}
\renewcommand{\thefootnote}{\fnsymbol{footnote}}
\newcounter{aff}
\renewcommand{\theaff}{\fnsymbol{aff}}
\newcommand{\affiliation}[1]{
\setcounter{aff}{#1} $\rule{0em}{1.2ex}^\theaff\hspace{-.4em}$}

%
\papertitlepage
\setcounter{page}{0}
\preprinumber{YITP-13-80}{} 
\vskip 2ex
\vfill
\begin{center}
{\large\bf\mathversion{bold}
Integrability of BPS equations in ABJM theory
}
\end{center}
\vfill
\baselineskip=3.5ex
\begin{center}
Kazuhiro Sakai\footnote[2]{\tt ksakai@fc.ritsumei.ac.jp}
and
Seiji Terashima\footnote[3]{\tt terasima@yukawa.kyoto-u.ac.jp}\\

{\small
\vskip 6ex
\affiliation{2}
{\it Department of Physical Sciences,
  Ritsumeikan University}\\[1ex]
{\it Shiga 525-8577, Japan}\\

\vskip 2ex
\affiliation{3}
{\it Yukawa Institute for Theoretical Physics, Kyoto University}\\[1ex]
{\it Kyoto 606-8502, Japan}

}
\end{center}
\vfill
\baselineskip=3.5ex
\begin{center} {\bf Abstract} \end{center}

We investigate BPS equations
which determine the configuration of
an M2--M5 bound state preserving half of the supersymmetries
in the ABJM theory.
We argue that the BPS equations are classically integrable,
showing that they admit a Lax representation.
The integrable structure of the BPS equations is
closely related to that of the Nahm equations.
Using this relation we formulate an efficient way
of constructing solutions of the BPS equations
from those of the Nahm equations.
As an illustration of our method,
we construct explicitly the most general solutions
describing two M2-branes suspended between two parallel M5-branes
as well as two semi-infinite M2-branes ending on an M5-brane.
These include previously unknown new solutions.
We also discuss a reduction of the BPS equations
in connection with the periodic Toda chain.

\vfill
\noindent
August 2013


\setcounter{page}{0}
\newpage
\renewcommand{\thefootnote}{\arabic{footnote}}
\setcounter{footnote}{0}
\setcounter{section}{0}
\baselineskip = 3.5ex
\pagestyle{plain}
%

\section{Introduction}

Towards understanding the still mysterious M-theory,
the theories on M2-branes and M5-branes
are expected to play crucial roles.
Concerning M2-branes, we now have
a strong candidate for the low energy effective theory on them,
i.e.~the Aharony--Bergman--Jafferis--Maldacena (ABJM) theory
\cite{Aharony:2008ug}.
On the other hand, the low energy effective theory on
multiple M5-branes is still unknown.
The theory should be self-dual 
under a certain ``electric--magnetic'' duality, which may mean that
construction of its action is unattainable
by conventional field theory techniques.

A possible way to investigate M5-branes
is to study M2--M5 bound states by means of the ABJM theory.
In doing this one can circumvent the above difficulty
because the ABJM action for $N$ M2-branes 
is an ordinary gauge field theory action with gauge group
$\grp{U}(N) \times \grp{U}(N)$.
Indeed, (classical) solutions 
describing M2--M5 bound states in the ABJM theory
have been found 
\cite{Terashima:2008sy,Gomis:2008vc,Terashima:2009fy,Terashima:2010ji}.
In particular, BPS equations
obtained in \cite{Terashima:2008sy,Gomis:2008vc} are of fundamental
importance. The equations can be considered as generalizations of 
the Basu--Harvey equations \cite{Basu:2004ed}
and determine the configuration of
an M2--M5 bound state preserving half of the supersymmetries
in the ABJM theory.
Explicit solutions of the BPS equations
have been constructed
\cite{Terashima:2008sy,Hanaki:2008cu,Nogradi:2005yk,Nosaka:2012tq}.

It has been known \cite{Diaconescu:1996rk} that
BPS conditions for certain D2--D4 bound states 
are described by the Nahm equations.
Because the M2--M5 bound state in M-theory is a counterpart
of the D2--D4 bound state in string theory, the BPS equations
in the ABJM theory
should have a close connection with the Nahm equations.
Indeed, it was found in \cite{Nosaka:2012tq} that
one can always construct two sets of solutions of the Nahm equations
from those of the BPS equations.\footnote{
This kind of relation was first found in
\cite{Gustavsson:2008dy} 
for the Bagger--Lambert--Gustavsson theory
\cite{Bagger:2007vi,Bagger:2007jr,Bagger:2006sk,Gustavsson:2007vu}.}
The Nahm equations are classically integrable \cite{Hitchin:1983ay}.
Moreover, based on the integrability
one can systematically construct
the Nahm data, i.e.~the solutions of the Nahm equations
\cite{Ercolani:1989tp,Manton:2004tk}.
It is natural to expect that the present BPS equations
are also integrable and one may be able to construct solutions
systematically making use of the integrability.

In this paper we argue that the BPS equations
in the ABJM theory are indeed classically integrable.
We find a Lax representation for the BPS equations.\footnote{
In \cite{Harland:2012cj},
it was argued that the Basu--Harvey equation is integrable and
a 3-bracket analog of the Lax representation
was presented. (We would like to thank C.~S\"amann for informing us
of the result.)}
As is expected, the integrable structure of the BPS equations is
closely related to that of the Nahm equations.
Using this relation we formulate an efficient way
of constructing solutions of the BPS equations
from those of the Nahm equations.
As an illustration,
we construct explicitly the most general solutions
describing two M2-branes suspended between two parallel M5-branes
as well as two semi-infinite M2-branes ending on an M5-brane.
These include previously unknown new solutions.

The organization of this paper is as follows.
In section~2, we present the Lax representation
of the BPS equations and introduce the auxiliary linear problem.
We then clarify the relation between the integrable structure
of the BPS equations and that of the Nahm equations.
Using this relation
we formulate a construction of solutions of the BPS equations
starting from those of the Nahm equations.
In section~3, we first study the auxiliary linear problem
of the simple funnel-type solution. We then
construct explicitly the most general solution
describing two semi-infinite M2-branes ending on an M5-brane,
following the method described in section~2.
We also construct the most general solution with $N=2$.
This solution is expressed in terms of elliptic functions
and describes two M2-branes suspended between two parallel M5-branes.
In section~4, we discuss a reduction of the BPS equations
in connection with the periodic Toda chain.
Appendix~A is a brief summary of the construction of the most general
Nahm data with $N=2$.
Conventions of special functions and some useful formulas
are summarized in Appendix~B.

\section{Integrability of BPS equations}

\subsection{BPS equations and Lax representation}

We are interested in BPS equations
in the ABJM theory and their solutions.
The equations are given by ordinary differential equations of
the following form\footnote{
The r.h.s.~of these equations differ from
those in \cite{Nosaka:2012tq} by an overall sign.
Correspondingly, the sign of the M5-brane charge of the solutions 
in this paper is opposite to that in \cite{Nosaka:2012tq}.
There are also several differences of conventions,
which we will not mention hereafter.}
\cite{Terashima:2008sy,Gomis:2008vc}
\[\label{BPSeqs}
\dot{Y}^a = Y^b Y^{b\dagger}Y^a - Y^aY^{b\dagger}Y^b.
\]
Here $Y^a(s)\ (a=1,2)$ are $N\times N$ complex matrices
and are functions in a real variable~$s$.
In this paper, a dot over letters stands for
the derivative in~$s$.
Note that the above equations have an
$\grp{SU}(N)\times\grp{SU}(N)\times \grp{SU}(2)\times \grp{U}(1)$
automorphism: If one transforms $Y^a$ as
\[\label{globalsym}
Y^a\to {Y'}^a= e^{i\varphi}\Lambda^a{}_b U Y^b V^\dagger
\]
with
\[
\qquad U,V\in\grp{SU}(N),\quad(\Lambda^a{}_b)\in\grp{SU}(2),
\quad e^{i\varphi}\in\grp{U}(1),
\]
${Y'}^a$ again satisfy the same equations as (\ref{BPSeqs}).

In this paper we argue that the BPS equations (\ref{BPSeqs})
are classically integrable.\footnote{
For the theory of classical integrable systems
the reader should refer to textbooks, for example,
\cite{Novikov:1984id,Babelon:2003}.}
We find that the equations can be expressed
in the form of the Lax equation:
\[\label{Laxeq}
\dot{A} = [A,B],
\]
where
\eqb
\label{LaxA}
A(s;\lambda)
\Eqn{=}\left(\begin{array}{cc}
O&Y^1+\lambda Y^2\\
Y^{1\dagger}-\lambda^{-1}Y^{2\dagger}&O
\end{array}\right),\\
\label{LaxB}
B(s;\lambda)
\Eqn{=}\left(\begin{array}{cc}
\lambda^{-1}Y^1 Y^{2\dagger}+\lambda Y^2 Y^{1\dagger}&O\\
O&\lambda Y^{1\dagger}Y^2+\lambda^{-1} Y^{2\dagger} Y^1
\end{array}\right).
\eqe
Here $\lambda\in\bbC$ is the spectral parameter.
There are actually various choice of Lax representations.\footnote{
For instance, one can take
\[
A'=
\left(\begin{array}{cc}
f(\lambda)\bvec{1}_N & O\\
O & g(\lambda)\bvec{1}_N
\end{array}\right)A,
\qquad
B'=B+h(A)
\]
with any nonzero functions $f,g,h$ as Lax operators.
Such $A'$ and $B'$
reproduce the same BPS equations.}
We have chosen the above representation so that
Lax operators admit several useful properties:
The above Lax operators manifest two kinds of
involution structures.
First, they admit the following $\bbZ_2$-parity structure
\[\label{Gammainvol}
\{A,\Gamma\}=0,\qquad[B,\Gamma]=0,
\]
where
\[
 \Gamma :=
\left(\begin{array}{cc}
\bvec{1}_N&0\\
0&- \bvec{1}_N
\end{array}\right).
\]
To see the other involution,
let us introduce the following operation
\[
{\cal M}^\star(\lambda):= {\cal M}(-\bar{\lambda}^{-1})^\dagger,
\]
which we will call a star-involution.
Here ${\cal M}$ is a matrix of any size and is a function in $\lambda$.
Under this involution, the operators $A,B$ transform as
\[\label{tildeconjLax}
A^\star=A,\qquad B^\star=-B.
\]
Another useful property is the following relation
\[\label{ABrelation}
B=\lambda\frac{\partial}{\partial\lambda}A^2. 
\]
We will make use of these structures to construct solutions.

As a general property of the Lax representation,
it follows from (\ref{Laxeq}) that
the eigenvalues of ${A}$ are independent of $s$.
Equivalently, the characteristic polynomial
\begin{align}
P:=&\,\det(\eta\bvec{1}_{2N}-{A})\nn\\
=&\,
  \det\left[\eta^2\bvec{1}_N-\left(Y^1+\lambda Y^2\right)
    \left(Y^{1\dagger}-\lambda^{-1}Y^{2\dagger}\right)\right]\nn\\
=&\,
  \det\left[
    \eta^2\bvec{1}_N-\left(Y^{1\dagger}-\lambda^{-1}Y^{2\dagger}\right)
    \left(Y^1+\lambda Y^2\right)\right]\qquad(\eta\in\bbC)
\end{align}
is independent of $s$.
It is convenient to introduce the notation
\[
\mu:=\eta^2.
\]
The characteristic equation
\[
P(\mu,\lambda)=0
\]
gives the spectral curve of a given solution.

\subsection{Auxiliary linear problem}

The Lax equation is regarded as the compatibility condition
of the following auxiliary linear problem
\eqb
\label{Lpsi}
{A}(s;\lambda)\psi(s;\lambda)\Eqn{=}\eta(\lambda)\psi(s;\lambda),\\
\label{Apsi}
{B}(s;\lambda)\psi(s;\lambda)\Eqn{=}-\dot{\psi}(s;\lambda).
\eqe
The first equation describes
an eigenvalue problem of the operator $A$,
while the second one describes the evolution in $s$ of the eigenvector
$\psi(s;\lambda)$.
Below we consider the case where
the operator $A$ possesses $2N$ linearly independent eigenvectors
$\psi_1,\ldots,\psi_{2N}$ with eigenvalues $\eta_1,\ldots,\eta_{2N}$.
The involution structure (\ref{Gammainvol})
implies that $A(\Gamma\psi_m)=-\eta_m(\Gamma\psi_m)$,
hence one can always choose the eigenvectors as
\[
\psi_{N+m}=\Gamma\psi_m,\qquad \eta_{N+m}=-\eta_m,\qquad m=1,\ldots,N.
\]
If we introduce the notation
\eqb
\label{MatrixPsi}
\iPsi\Eqn{:=}(\psi_1,\ldots,\psi_{2N})
=(\psi_1,\ldots,\psi_N,\Gamma\psi_1,\ldots,\Gamma\psi_N),\\
\label{Matrixeta}
D\Eqn{:=}\diag(\eta_1,\ldots,\eta_{2N})
=\diag(\eta_1,\ldots,\eta_N,-\eta_1,\ldots,-\eta_N),
\eqe
the auxiliary linear problem can be expressed as
\eqb
\label{APsi}
A\iPsi \Eqn{=} \iPsi D,\\
\label{BPsi}
B\iPsi \Eqn{=} -\dot{\iPsi}.
\eqe
Since we have assumed that
there exist $2N$ linearly independent eigenvectors,
$\iPsi$ is a regular matrix at generic values of $s$ and $\lambda$.
Therefore, one can invert (\ref{APsi}) and express the operator $A$
as
\[\label{AinPsi}
A(s;\lambda)=\iPsi(s;\lambda) C(\lambda)\iPsi^\star(s;\lambda),
\]
where
\eqb
C\Eqn{:=}D{\cal N}^{-1},\\
{\cal N}\Eqn{:=}\iPsi^\star \iPsi.
\eqe
It is easy to verify that ${\cal N}$ and $C$
are independent of $s$ and are self-adjoint with respect to
the star-involution.

The above structure implies that
a generic solution of the original BPS equations can be expressed
as a bilinear combination of $\iPsi(s;\lambda)$.
As is well known, for a system with Lax representation
there are several powerful techniques to
restrict the form of $\iPsi(s;\lambda)$ and construct a class of
general solutions.
In the present case the situation is even better,
due to the fact that the BPS equations are
closely related to the Nahm equations.
Solutions of the Nahm equations are well studied
and using them one can easily determine the eigenvectors
$\iPsi(s;\lambda)$, as we will explain in the next subsection.

\subsection{Relation to Nahm equations and construction of solutions}

Recall that the Nahm equations are given as
\[
 \dot{T}^I = i \epsilon_{IJK} T^J T^K.
\label{Nahmeq}
\]
Here indices $I,J,K$ take values $1,2,3$ and
$T^I$ are $N \times N$ hermitian matrices.
It was found in \cite{Nosaka:2012tq} that
one can always construct two sets of solutions of the Nahm equations
from those of the BPS equations (\ref{BPSeqs}). Indeed,
if $Y^a$ are solutions of (\ref{BPSeqs}), bilinear combinations
\begin{eqnarray}
\label{TinY}
T^I_1 := (\sigma^I)_{ab} Y^a Y^{b\dagger},\qquad
T^I_2 := (\sigma^I)_{ab} Y^{b\dagger} Y^a
\label{Nahmdata}
\end{eqnarray}
both satisfy the Nahm equations (\ref{Nahmeq}).
Here $\sigma^I$ denote Pauli matrices.

It has been known that the Nahm equations are classically integrable
and admit a Lax representation \cite{Hitchin:1983ay}.
Lax equations for the above
$T^I_\alpha\ (\alpha=1,2)$ are written
as
\[\label{subLaxeq}
 \dot{A}_\alpha =[A_\alpha,B_\alpha],
\]
where
\begin{eqnarray}
A_\alpha \Eqn{:=} T_\alpha^3 +\frac{\lambda}{2}
  \left(T_\alpha^1 - i T_\alpha^2 \right)
-\frac{1}{2 \lambda} \left(T_\alpha^1 + i T_\alpha^2 \right), \\
B_\alpha \Eqn{:=}
\frac{\lambda}{2} \left(T_\alpha^1 - i T_\alpha^2 \right)
+\frac{1}{2 \lambda} \left(T_\alpha^1 + i T_\alpha^2 \right).
\end{eqnarray}

As the reader may expect,
these Lax operators are directly related to those
for the BPS equations (\ref{LaxA}), (\ref{LaxB}).
They are related in a remarkably simple way:
\begin{eqnarray}
\label{Laxrel}
 A^2 =
\left(\begin{array}{cc}
A_1 &0\\
0& A_2
\end{array}\right),\qquad
 B =
\left(\begin{array}{cc}
B_1 &0\\
0& B_2
\end{array}\right).
\end{eqnarray}
This relation suggests that
the Lax operator $A$ for the BPS equations
is regarded as the ``square root'' of that for the Nahm equations.
This is somewhat
analogous to the relation between the Dirac operator
and the Laplace operator.

The above relation implies an important fact:
Any eigenvector of $A$ is divided into
an eigenvector of $A_1$ and that of $A_2$
with a common eigenvalue.
This fact provides us with an efficient way of
constructing solutions of the BPS equations,
as we explain below.

Suppose that we are given a pair of Nahm data
$T_\alpha^I$ for which the Lax operators
$A_\alpha\ (\alpha=1,2)$
have the same set of eigenvalues
$\mu_1(\lambda),\ldots,\mu_N(\lambda)$.
Put differently, the pair of Nahm data share
the same spectral curve
\[
P(\mu,\lambda)=0
\]
with
\[
P=\det(\mu\bvec{1}_N-A_1)=\det(\mu\bvec{1}_N-A_2).
\]
We assume that each of $A_\alpha$
has $N$ linearly independent eigenvectors.
In this case, the eigenvalue problems of $A_\alpha$ can be expressed
as
\[\label{APsialpha}
 A_\alpha \iPsi_\alpha = \iPsi_\alpha M
\]
with
\[
M = \diag(\mu_1,\ldots,\mu_N).
\]
Here $\iPsi_\alpha(s;\lambda)$ are regular matrices
at generic values of $s$ and $\lambda$.
Since the Lax equations (\ref{subLaxeq}) hold,
one can always normalize $\iPsi_\alpha$ in such a way that
they satisfy
\[\label{BPsialpha}
B_\alpha\iPsi_\alpha = -\dot{\iPsi}_\alpha.
\]
By using this equation and $B_\alpha\pres^\star=-B_\alpha$,
one can verify that
\[
{\cal N}_\alpha:=\iPsi_\alpha\pres^\star\iPsi_\alpha
\]
are independent of $s$.

As we mentioned above, if we express a solution (\ref{MatrixPsi})
of the auxiliary linear problem (\ref{APsi}), (\ref{BPsi}) as
\[\label{PsiinPsialpha}
\iPsi=
\frac{1}{\sqrt{2}}
\left(\begin{array}{cc}
\iPsi_1&\iPsi_1 \\
\iPsi_2&-\iPsi_2
      \end{array}\right),
\]
the submatrices $\iPsi_\alpha$ are automatically
solutions of the equations (\ref{APsialpha}), (\ref{BPsialpha})
with $A_\alpha,B_\alpha$ given by (\ref{Laxrel}).
On the other hand, if we start from
a given pair of Nahm data (sharing the same spectral curve)
and put eigenvectors $\iPsi_\alpha$ into the form
(\ref{PsiinPsialpha}), it does not always give a solution
of the auxiliary linear problem of the BPS equations.
Below let us examine under what conditions
eigenvectors $\iPsi_\alpha$ for given Nahm data
generate a solution of the BPS equations.

The above $\iPsi$ is associated with
the eigenvalue matrix (\ref{Matrixeta}) of the form
\[
D=
\left(\begin{array}{cc}
H&O\\
O&-H
      \end{array}\right),
\qquad H=\diag(\eta_1,\ldots,\eta_N).
\]
Since operators $A$ and $A_\alpha$ are related as in (\ref{Laxrel}),
their eigenvalues are related as
\[
H^2=M.
\]
As we saw in the last subsection,
the operator $A$ can be constructed from $\iPsi$ as in (\ref{AinPsi}).
One could start from this general form, but here
it is not difficult to guess the form of $A$:
One can check that
the above $\iPsi$ indeed gives the eigenvectors of the eigenvalue
problem (\ref{APsi}) if the operator $A$ is given as
\[
\label{AinPsi_alpha}
A=
\left(\begin{array}{cc}
O&\iPsi_1 H {\cal N}_2^{-1}\iPsi_2^\star \\
\iPsi_2 H {\cal N}_1^{-1}\iPsi_1^\star&O
      \end{array}\right).
\]
In order for $A$ to have the property $A^\star=A$,
it is required that
\[\label{HNrel}
H{\cal N}_1={\cal N}_2 H.
\] 
The evolution equation (\ref{BPsi})
automatically follows from (\ref{BPsialpha}).
The remaining requirement
is that the upper-right block of (\ref{AinPsi_alpha})
is linear in $\lambda$, i.e.
\[\label{linearity}
\frac{\partial^2}{\partial\lambda^2}
\left[\iPsi_1 H{\cal N}_2^{-1}\iPsi_2^\star\right]=0.
\]
If both of the above conditions are satisfied,
one obtains a consistent realization of the auxiliary linear problem
of the BPS equations.
The solutions $Y^a$ of the original BPS equations
are then obtained explicitly as
\[\label{YfromA12}
Y^1=\iPsi_1 H{\cal N}_2^{-1}\iPsi_2^\star\bigg|_{\lambda=0},
\qquad
Y^2=\frac{\partial}{\partial\lambda}
\left[\iPsi_1 H{\cal N}_2^{-1}\iPsi_2^\star\right]\bigg|_{\lambda=0}.
\]

To sum up, we have seen that
the solutions of the BPS equations can be constructed
starting from the solutions of the Nahm equations.
The construction of Nahm data has been extensively studied
\cite{Ercolani:1989tp, Manton:2004tk}.
Given a pair of Nahm data $T_\alpha^I$, it is in principle
a straightforward task to compute eigenvectors $\iPsi_\alpha$.
The condition (\ref{HNrel}) can be easily satisfied:
For instance, it is satisfied if we normalize $\iPsi_\alpha$ as
\[
{\cal N}_1={\cal N}_2=\diag(n_1(\lambda),\ldots,n_N(\lambda))
\]
with some nonzero functions $n_m(\lambda)$.
On the other hand, the condition (\ref{linearity})
usually gives nontrivial constraints.
To the best of our knowledge,
we have to deal with this condition case by case.
Nevertheless, the above procedure provides us with an efficient
way of constructing a wide class of solutions of the BPS equations.
In the next section, we will illustrate this construction
by explicit examples.

\section{General solutions}

In this section we construct new solutions of the BPS equations
following the method described in the last section.
We first discuss the funnel-type solution with general $N$
and then focus on the general solutions with $N=2$.
The solutions with $N=2$ are of fundamental importance,
since a wide class of solutions with $N>2$ written in terms of
elementary functions or elliptic functions are, up to
the global symmetry transformation (\ref{globalsym}),
direct sums of these solutions.

General solutions with $N=2$ are written in terms of elliptic
functions. For the purpose of illustrating our method, however,
it is enough to work on the semi-infinite solutions,
which are written simply in terms of hyperbolic functions.
We then generalize them to elliptic solutions and see that
they indeed correspond to the most general Nahm data with $N=2$.

\subsection{Funnel-type solution and auxiliary linear problem}

Let us first consider the funnel-type solution \cite{Nosaka:2012tq}.
In this case we already know the solution, but
as a warm up exercise
let us verify that the solution is indeed derived from
the corresponding Nahm data.

The solution is given as
\[\label{funnelsol}
Y^1=\sqrt{\frac{c}{1-e^{-2x}}}G^1,\qquad
Y^2=\sqrt{\frac{c}{e^{2x}-1}}G^2,
\]
where
\[
x=c s,\qquad c\ge 0
\]
and $G^a$ are constant $N\times N$ matrices satisfying
\[\label{Grel}
-G^a = G^b G^{b\dagger}G^a - G^aG^{b\dagger}G^b.
\]
One can take $G^a$ as
\[
(G^1)_{mn}=\sqrt{N-n}\,\delta_{m,n},\qquad
(G^2)_{mn}=\sqrt{n}\,\delta_{m,n+1}.
\]
Let us next introduce
\[
\label{tauinY}
\tau^i_1 := \frac{1}{2}(\sigma^i)_{ab} G^a G^{b\dagger},\qquad
\tau^i_2 := \frac{1}{2}(\sigma^i)_{ab} G^{b\dagger} G^a\qquad
(i=1,2,3,4),
\]
where $\sigma^I\ (I=1,2,3)$ are Pauli matrices and
$\sigma^4=\bvec{1}_2$. It follows from (\ref{Grel}) that
\[\label{taualg}
[\tau_\alpha^I,\tau_\alpha^J]=i\epsilon_{IJK}\tau_\alpha^K,\qquad
[\tau_\alpha^I,\tau_\alpha^4]=0.
\]
In fact, $\tau_1^I$ and $\tau_2^I$ $(I=1,2,3)$ are
$\bvec{N}$ and $\bvec{N-1}\oplus\bvec{1}$
representations of $\grp{SU}(2)$, respectively.
By using the relation (\ref{TinY}),
the corresponding Nahm data are obtained as
\[
T_\alpha^1=\frac{c}{\sinh x}\tau_\alpha^1,\qquad 
T_\alpha^2=\frac{c}{\sinh x}\tau_\alpha^2,\qquad 
T_\alpha^3=\frac{c}{\tanh x}\tau_\alpha^3
  +c\tau_\alpha^4.
\]

Let us now regard the above Nahm data as input
and derive the solution of the BPS equation
following the method explained in the last section.
Here we first assume that $c>0$ and
later comment on the limit $c\to 0$.
The Lax operators for the Nahm data are written as
\[
A_\alpha=c\left(
\frac{1}{\tanh x}\rho_\alpha^3+\rho_\alpha^4
-\frac{i}{\sinh x}\rho_\alpha^2\right),\qquad
B_\alpha=\frac{c}{\sinh x}\rho_\alpha^1,
\]
where
\begin{align}
\rho_\alpha^1&:=\frac{\lambda+\lambda^{-1}}{2}\tau_\alpha^1
  +\frac{\lambda-\lambda^{-1}}{2i}\tau_\alpha^2,\qquad&
\rho_\alpha^2&:=-\frac{\lambda-\lambda^{-1}}{2i}\tau_\alpha^1
  +\frac{\lambda+\lambda^{-1}}{2}\tau_\alpha^2,\qquad\nn\\
\rho_\alpha^3&:=\tau_\alpha^3,\qquad&
\rho_\alpha^4&:=\tau_\alpha^4.
\end{align}
Clearly, $\rho_\alpha^i$ satisfy the same algebra (\ref{taualg})
as $\tau_\alpha^i$ do.
Using this fact,
one can easily diagonalize $A_\alpha$ as
\[
A_\alpha
   =c\left(\tanh\frac{x}{2}\right)^{-\rho_\alpha^1}
     \left(\rho_\alpha^3+\rho_\alpha^4\right)
     \left(\tanh\frac{x}{2}\right)^{\rho_\alpha^1}.
\]
One can check that
\[
\iPsi_\alpha=\left(\tanh\frac{x}{2}\right)^{-\rho_\alpha^1}
\]
is a solution of the auxiliary linear problem
(\ref{APsialpha}), (\ref{BPsialpha})
with
\[
M=c\left(\rho_\alpha^3+\rho_\alpha^4\right)=c\left(G^1\right)^2.
\]
Observing that $(\rho_\alpha^1)^\star=-\rho_\alpha^1$,
one finds that
\[
{\cal N}_\alpha = \bvec{1}_N.
\]
One can take $H=M^{1/2}$ as
\[
H=\sqrt{c}G^1.
\]
Substituting these into the general expression (\ref{AinPsi_alpha}),
one obtains the explicit form of the Lax operator $A$.
The results indeed takes the form (\ref{LaxA})
with the original solution (\ref{funnelsol}).

In the limit $c\to 0$,
the auxiliary linear problem becomes degenerate,
i.e.~the Lax operator $A$ does not have $2N$ independent
eigenvectors. Nevertheless, one finds that (\ref{funnelsol})
does give a well-defined solution $Y^a=(2s)^{-1/2}G^a$
in the limit $c\to 0$.
In this way, it is often possible to obtain a solution
even if the corresponding auxiliary linear problem is degenerate.

\subsection{Semi-infinite solutions}

In this subsection we construct the general BPS solution
describing two semi-infinite M2-branes ending on an M5-brane.
We set the M5-brane at $s=0$ and consider the solution
over the semi-infinite line $s>0$.
The solutions $Y^a(s)$ are regular for $s>0$ and diverge at $s=0$.
For $s\to\infty$, they
approach constant values, i.e.~
\[
\dot{Y}^a(\infty)=0.
\]
Using the BPS equations at $s=\infty$, one can check that
$Y^1(\infty)$ and $Y^2(\infty)$ can be simultaneously diagonalized
by a suitable transformation $Y^a\to UY^aV^\dagger$
with $U,V\in\grp{SU}(2)$. 
Note that the values of diagonal elements of $Y^a(\infty)$
represent the locations of two M2-branes in the four directions
tangent to the M5-brane. 

When both of the matrices
$Y^1(\infty)$ and $Y^2(\infty)$ are diagonal,
the corresponding Nahm data $T_\alpha^I(\infty)$ are
also diagonal and
\[\label{T1T2atinfty}
T_1^I(\infty)=T_2^I(\infty).
\]
It is shown in Appendix~A that any Nahm data with $N=2$ can be
expressed in the canonical form (\ref{Nahmgenform})
by a suitable transformation.
Nahm data of the form (\ref{Nahmgenform})
which satisfy the above properties at $s=\infty$
are easily determined as\footnote{
To be precise, only one of the two sets of Nahm data
can be transformed into the canonical form (\ref{Nahmgenform}),
because the $\grp{SO}(3)$ automorphism
$T_\alpha^I\to L^I_J T_\alpha^J$ acts simultaneously on
both $T_1^I$ and $T_2^I$.
If we transform $T_1^I$ into the canonical
form, the coefficients of Pauli matrices in $T_2^I$
are left with $\grp{SO}(3)$ degrees of freedom a priori.
In the present case, however,
only the 1-2 rotation of the $\grp{SO}(3)$ is compatible with
the boundary condition (\ref{T1T2atinfty}).
Furthermore, this degree of freedom can be removed by the
$\grp{SU}(2)$ transformation $T_2^I\to VT_2^IV^\dagger$
because in the present case
all the offdiagonal components of $T_2^1$ and $T_2^2$
are proportional to a single function $c/\sinh(x-x_2)$.
As a result, one can take the two sets of Nahm data as
in (\ref{semiinfNahm}) without loss of generality.
}
\eqb
\label{semiinfNahm}
T_\alpha^1\Eqn{=}\frac{c}{\sinh(x-x_\alpha)}\frac{\sigma^1}{2}
  +t^1\bvec{1}_2,\qquad
T_\alpha^2=\frac{c}{\sinh(x-x_\alpha)}\frac{\sigma^2}{2}
  +t^2\bvec{1}_2,\nn\\
T_\alpha^3\Eqn{=}\frac{c}{\tanh(x-x_\alpha)}\frac{\sigma^3}{2}
  +t^3\bvec{1}_2.
\eqe
Here $t^I,\,x_\alpha\ (\alpha=1,2)$ are real constants and
\[
x=cs,\qquad c\ge 0.
\]
Since the solutions $Y^a(s)$ are regular for $s>0$, both sets of
Nahm data $T_\alpha^I$ have to be regular at least for $s>0$.
The singular behavior of $Y^a(s)$ at $s=0$ implies that
at least one set of the Nahm data are also singular at $s=0$.
To satisfy these conditions, one can take
\[\label{x1x2inl}
x_1=0, \qquad x_2=-l,\qquad l\ge 0,
\]
without loss of generality.
Note that one can take $T^I_2$ as constant solutions.
This is realized as a special case where $l$ is sent to infinity.

The Lax operators are obtained as
\eqb
A_\alpha
 \Eqn{=}\left(\tanh\frac{x-x_\alpha}{2}\right)^{-\rho^1}
  M \left(\tanh\frac{x-x_\alpha}{2}\right)^{\rho^1}\\
\noalign{with}
M\Eqn{=}\left(\frac{c}{2}\sigma^3+t_\lambda\bvec{1}_2\right),\\
\rho^1\Eqn{=}\frac{\lambda+\lambda^{-1}}{4}\sigma^1
  +\frac{\lambda-\lambda^{-1}}{4i}\sigma^2
=\frac{1}{2}\left(\begin{array}{cc}
 0 & \lambda^{-1} \\ \lambda & 0
       \end{array}\right),\\
t_\lambda
  \Eqn{=}t^3+\frac{\lambda}{2}(t^1-it^2)-\frac{1}{2\lambda}(t^1+it^2).
\eqe
The eigenvector matrices of $A_\alpha$ are
given as
\[
\iPsi_\alpha=\left(\tanh\frac{x-x_\alpha}{2}\right)^{-\rho^1}
D_\alpha,
\]
where $D_\alpha$ are diagonal matrices
satisfying
\[
D_\alpha\pres^\star D_\alpha = \bvec{1}_2.
\]
The above $\iPsi_\alpha$ are normalized as
\[
{\cal N}_\alpha = \bvec{1}_2.
\]

We are now in a position to
impose the condition (\ref{linearity}).
To do this, let us first evaluate the matrix
\eqb
\label{A12rational}
\iPsi_1 H{\cal N}_2^{-1}\iPsi_2^\star
\Eqn{=}\left(\tanh\frac{x-x_1}{2}\right)^{-\rho^1}
  D_1M^{1/2}D_2^\star\left(\tanh\frac{x-x_2}{2}\right)^{\rho^1}.
\eqe
The matrix $D_1M^{1/2}D_2^\star$ is diagonal.
We express it as
\[
D_{12}:=D_1M^{1/2}D_2^\star
=\left(\begin{array}{cc}
 \delta_+(\lambda)&0 \\[1ex] 0&\delta_-(\lambda)
       \end{array}\right).
\]
By observing
\eqb
\left(\tanh\frac{x}{2}\right)^{\pm \rho^1}
\Eqn{=}\frac{1}{\sqrt{2\sinh x}}
  \left(\begin{array}{cc}
 e^{x/2}& \mp\lambda^{-1}e^{-x/2}\\[1ex]
 \mp\lambda e^{-x/2} & e^{x/2}
       \end{array}\right)
\eqe
and using the values of $x_\alpha$ specified in (\ref{x1x2inl}),
the matrix (\ref{A12rational}) is evaluated as
\eqb
\lefteqn{\iPsi_1 H{\cal N}_2^{-1}\iPsi_2^\star}\nn\\
\Eqn{=}\frac{1}{2\sqrt{\sinh x\sinh(x+l)}}
  \left(\begin{array}{cc}
  \delta_+ e^{x+l/2}-\delta_- e^{-x-l/2}&
  \lambda^{-1}\left(-\delta_+ e^{-l/2}+\delta_- e^{l/2}\right)\\[2ex]
  \lambda\left(\delta_+ e^{l/2}-\delta_- e^{-l/2}\right)&
  -\delta_+ e^{-x-l/2}+\delta_- e^{x+l/2}
	\end{array}\right).\qquad\ 
\eqe
In order for the diagonal components of this matrix
to be linear in $\lambda$, $\delta_\pm(\lambda)$
have to be linear in $\lambda$.
Moreover, in order for the off-diagonal components also to be
linear in $\lambda$, $\delta_\pm(\lambda)$ have to satisfy
\[
\delta_+ e^{l/2}-\delta_- e^{-l/2}=\alpha,\qquad
-\delta_+ e^{-l/2}+\delta_- e^{l/2}=\beta\lambda
\]
with some constants $\alpha, \beta\in\bbC$.
By solving these constraints, $\delta_\pm$ are expressed as
\[\label{deltaexp}
\delta_\pm(\lambda) = 
  \frac{e^{\pm l/2}\alpha+e^{\mp l/2}\beta\lambda}
       {e^l-e^{-l}}.
\]
Next, observe that
\[
D_{12}\pres^\star D_{12}=M.
\]
Together with (\ref{deltaexp}), this condition implies that
\eqb
\alpha\bar{\alpha}\Eqn{=}
  \sinh l\left(c+2t^3\tanh l\right),\qquad
\alpha\bar{\beta}=
  2\sinh^2 l\,(t^1+it^2),\nn\\
\beta\bar{\beta}\Eqn{=}
  \sinh l\left(c-2t^3\tanh l\right).
\eqe
From the consistency of these equations,
parameters $l,c,t^I$ have to satisfy
\[
\left(t^1\right)^2+\left(t^2\right)^2
  +\frac{\left(t^3\right)^2}{\cosh^2l}
=\frac{c^2}{4\sinh^2 l}.
\]
This is solved as
\[
t^1=\frac{c}{2\sinh l}n_1,\qquad
t^2=\frac{c}{2\sinh l}n_2,\qquad
t^3=\frac{c}{2\tanh l}n_3
\]
with
$\bvec{n}=(n_1,n_2,n_3)^{\bf T}$ being a three-dimensional unit vector.
We parametrize it by
\[\label{nIinthetaphi}
(n_1,n_2,n_3)=
(\sin\theta\cos\phi,
\sin\theta\sin\phi,
\cos\theta)
\]
with
$0\le\theta\le\pi,\,0\le\phi<2\pi$.
By using this parametrization, $\alpha, \beta$ are solved as
\eqb
\alpha\Eqn{=}
  \sqrt{2c\sinh l}\cos\frac{\theta}{2}\,e^{i\chi+i\phi},\qquad
\beta=
  \sqrt{2c\sinh l}\sin\frac{\theta}{2}\,e^{i\chi}
\eqe
with $\chi\in\bbR$.
The common phase factor $e^{i\chi}$ will eventually become
an overall phase factor of the solution. It can be removed
by the $\grp{U}(1)$ rotation $Y^a\to e^{-i\chi}Y^a$.
Finally, using (\ref{YfromA12}) one obtains
\eqb
\label{semiinfY}
Y^1\Eqn{=}\sqrt{\frac{c}{2\sinh l\sinh x\sinh (x+l)}}
  \left(\begin{array}{cc}
   \sinh(x+l)\cos\frac{\theta}{2}\,e^{i\phi}&
   \sinh l\sin\frac{\theta}{2}\\[1ex]
   0&
   \sinh x\cos\frac{\theta}{2}\,e^{i\phi}
	\end{array}\right),\nn\\
Y^2\Eqn{=}\sqrt{\frac{c}{2\sinh l\sinh x\sinh (x+l)}}
  \left(\begin{array}{cc}
   \sinh x\sin\frac{\theta}{2}&
   0\\[1ex]
   \sinh l\cos\frac{\theta}{2}\,e^{i\phi}\ &
   \sinh (x+l)\sin\frac{\theta}{2}
	\end{array}\right).
\eqe

The funnel-type solution in the last subsection is obtained
by setting $\theta=\phi=0$ and taking the limit $l\to\infty$.

By construction this solution serves as the general solution
with $N=2$ satisfying the semi-infinite boundary condition,
up to translation in $x$
and the automorphism (\ref{globalsym}).
The solution possesses
four free parameters $c,l,\theta,\phi$.
The number of free parameters coincides with that of
the most general semi-infinite solution (\ref{semiinfNahm})
of the Nahm equations: The solution (\ref{semiinfNahm}) also
possesses four free parameters $c$ and $t^I$,
up to translation in $x$ and the $\grp{SU}(2)\times\grp{SO}(3)$
automorphism.
It is interesting that one of the two sets of Nahm data
$T_\alpha^I\ (\alpha=1,2)$ essentially determine
the form of $Y^a$ as well as the other set of Nahm data.
In particular, the distance of two singularities $l=x_1-x_2$
is uniquely determined if $c$ and $t^I$ are specified.
From the point of view of M-theory, the coincidence of
the degrees of freedom is regarded as the correspondence
of the numbers of moduli between
M2--M5 bound states and D2--D4 bound states.

The above number of free parameters is also consistent with
the moduli counting in the ABJM theory.
As we mentioned in the beginning of this subsection,
$Y^1(\infty)$ and $Y^2(\infty)$ are simultaneously diagonalized
by an $\grp{SU}(2)\times\grp{SU}(2)$ automorphism transformation
$Y^a\to UY^aV^\dagger$.
The values of diagonal elements of $Y^a(\infty)$
give eight real moduli, which represent the locations of
two M2-branes in the four directions
tangent to the M5-brane.
Four of the eight moduli correspond to
the remaining $\grp{SU}(2)\times \grp{U}(1)$ part of the
automorphism (\ref{globalsym}), i.e.~the global symmetry
of the solutions. The other four moduli are identified with
the above $c,l,\theta,\phi$ which
characterize the ``shape'' of the solutions.
Notice that in the ABJM theory the above $\grp{U}(1)$
has to be regarded as a part of moduli rather than
a gauge degree of freedom \cite{Aharony:2008ug}.

\subsection{General elliptic solutions}

The general semi-infinite solution (\ref{semiinfY})
constructed in the last subsection 
can be expressed as
\eqb
Y^1\Eqn{=}\frac{1}{2}\left(
  f_1\sin\frac{\theta}{2}\,\sigma^1
 +f_2\sin\frac{\theta}{2}\, i\sigma^2
 +f_3e^{i\phi}\cos\frac{\theta}{2}\,\sigma^3
 -f_0e^{i\phi}\cos\frac{\theta}{2}\,\bvec{1}_2\right),\nn\\
Y^2\Eqn{=}\frac{1}{2}\left(
  f_1e^{i\phi}\cos\frac{\theta}{2}\,\sigma^1
 -f_2e^{i\phi}\cos\frac{\theta}{2}\,i\sigma^2
 -f_3\sin\frac{\theta}{2}\,\sigma^3
 -f_0\sin\frac{\theta}{2}\,\bvec{1}_2\right),
\label{ellY12}
\eqe
where $\sigma^I$ are Pauli matrices and
\[
\label{fiforsemiinf}
f_1=f_2=\sqrt{\frac{c\sinh l}{2\sinh x\sinh(x+l)}},\quad\
f_3=\frac{\cosh (x+l/2)}{\cosh (l/2)}f_1,\quad\ 
f_0=-\frac{\sinh (x+l/2)}{\sinh (l/2)}f_1.
\]
As wee see below,
making use of this expression one can construct
the general solution of the BPS equations for $N=2$ without
repeating the procedure in the last subsection.

Let us first note that the above $f_i(s)$ are real functions
and satisfy differential equations
\[\label{fODE}
\dot{f}_i=f_jf_kf_l,
\]
where the values of $i,j,k,l$ are taken to be all distinct.
Interestingly, the differential equations (\ref{fODE})
(and the restriction that $f_i$ are real functions)
are sufficient for $Y^a(s)$ of the form (\ref{ellY12})
to satisfy the BPS equations (\ref{BPSeqs}).
This means that one immediately obtains more general solutions
of the BPS equations by merely finding 
more general solutions of (\ref{fODE}).
Observing that $f_i^2-f_j^2$ are constants,
One can reduce (\ref{fODE}) to a single first-order differential
equation and construct the general solution.
In fact, the general solution of (\ref{fODE})
has already been obtained in \cite{Nosaka:2012tq}.
In what follows, we present the solution in several new expressions.
These new expressions are certainly useful for a clear understanding
of the structure of the solution.

A sufficiently general solution of (\ref{fODE}), which contains four
independent parameters,
is given by
\begin{equation}
\label{fintheta}
f_i=\frac{\varth_{i+1}(u)}{\varth_{i+1}(u_\ast)}\sqrt{
\frac{\pi}{2\omega_1}
\frac{\varth_1(u_\ast)\varth_2(u_\ast)\varth_3(u_\ast)\varth_4(u_\ast)}
     {\varth_1(u_\ast+u)\varth_1(u_\ast-u)}}
\qquad (i=0,1,2,3).
\end{equation}
Here $\varth_{i+1}(u):=\varth_{i+1}(u,\tau)$ are Jacobi theta functions
(see Appendix~B) and the variable $u$ is defined as
\[
u=\frac{s-s_0}{2\omega_1}.
\]
This solution contains four parameters
\[
s_0\in\bbR,\qquad 0<u_\ast<\frac{1}{2},\qquad
\omega_1\in\bbR_{>0},\qquad \tau\in i\>\!\bbR_{>0}.
\]
The solution is defined over the region
\[
-u_\ast<u<u_\ast.
\]
At each boundary of this region $f_i$ diverge.
This implies that M2-branes are bounded by an M5-brane
located at each of these boundaries.
Thus, the present solution describes two M2-branes
suspended between two parallel M5-branes.

Just outside this region the above $f_i$ become
imaginary. However, slightly modifying them as
\begin{equation}
\label{finthetabis}
\tilde f_i=(-1)^{\delta_{1,i}}
\frac{\varth_{i+1}(u)}{\varth_{i+1}(u_\ast)}\sqrt{
\frac{\pi}{2\omega_1}
\frac{\varth_1(u_\ast)\varth_2(u_\ast)\varth_3(u_\ast)\varth_4(u_\ast)}
     {\varth_1(u+u_\ast)\varth_1(u-u_\ast)}}
\qquad (i=0,1,2,3),
\end{equation}
one again obtains a set of real functions which satisfy
the differential equations (\ref{fODE}).
This solution is now defined over the region
\[
u_\ast<u<1-u_\ast.
\]

The above $f_i$ and $\tilde{f}_i$ satisfy
\begin{align}
\label{finequality}
f_1 \ge f_2 \ge f_3 > -f_0 > 0&
\qquad\mbox{for}\qquad -u_\ast < u < 0,\nn\\
f_1 \ge f_2 \ge f_3 > f_0 > 0&
\qquad\mbox{for}\qquad 0 < u < u_\ast,\\
\label{fhatinequality}
\tilde{f}_0 \ge \tilde{f}_3 \ge \tilde{f}_2 >-\tilde{f}_1 > 0&
\qquad\mbox{for}\qquad u_\ast < u <\frac{1}{2},\nn\\
\tilde{f}_0 \ge \tilde{f}_3 \ge \tilde{f}_2 >\tilde{f}_1 > 0&
\qquad\mbox{for}\qquad \frac{1}{2}< u < 1-u_\ast.
\end{align}
Any permutation of $f_i$ and/or
overall sign change of 
even number of $f_i$ again give another solution
to the equations (\ref{fODE}), where the new $f_i$ satisfy
different inequalities from (\ref{finequality}). 
The same holds for $\tilde{f}_i$.
All these constitute
the general solution of (\ref{fODE}).

In the following,
let us concentrate on the solution (\ref{fintheta})
and study its properties.
The solution is also concisely expressed
in terms of Weierstrass elliptic functions (see Appendix~B) as
\begin{equation}
\label{finwp}
f_0=
\left(\frac{\wp_1(s_\ast)\wp_2(s_\ast)\wp_3(s_\ast)}
           {\wp(s-s_0)-\wp(s_\ast)}\right)^{1/2},\qquad
f_I=\frac{\wp_I(s-s_0)}{\wp_I(s_\ast)}f_0\quad(I=1,2,3),
\end{equation}
where
\[
s_\ast = 2\omega_1 u_\ast,\qquad 0<s_\ast<\omega_1.
\]
This solution is defined over the region $s_0-s_\ast<s<s_0+s_\ast$
and the branch of $f_0$ is taken in such a way that
$f_0\gtrless 0$ for $0\lessgtr s-s_0\lessgtr \pm s_\ast$.

As mentioned previously, $f_I^2-f_0^2$ are constants.
These constants are expressed as
\begin{align}
f_I^2-f_0^2
&=\frac{\pi\varth_{I+1}^2}{2\omega_1}
  \frac{\varth_{J+1}(u_\ast)\varth_{K+1}(u_\ast)}
       {\varth_{1}(u_\ast)\varth_{I+1}(u_\ast)}\nn\\
&=\frac{\wp_J(s_\ast)\wp_K(s_\ast)}{\wp_I(s_\ast)}\nn\\
&=:a_I^2
\qquad(a_I> 0),
\label{alphaIdef}
\end{align}
where $I,J,K$ are any permutation of $1,2,3$.
It can be checked that
\[
a_1\ge a_2\ge a_3> 0.
\]
Instead of $\omega_1,\tau,s_\ast, s_0$,
one could use $a_1,a_2,a_3,s_0$ as free parameters.
Using Jacobi elliptic functions,
one can express the solution (\ref{fintheta}) as
\begin{align}
\qquad
f_0&=
  \frac{a_3\sn x}{\sqrt{\sn^2 x_\ast-\sn^2 x}},\quad&
f_1&=
  \frac{a_1\sn x_\ast\cn x}{\sqrt{\sn^2 x_\ast-\sn^2 x}},\nn\\
f_2&=
  \frac{a_2\sn x_\ast\dn x}{\sqrt{\sn^2 x_\ast-\sn^2 x}},\quad&
f_3&=
  \frac{a_3\sn x_\ast}{\sqrt{\sn^2 x_\ast-\sn^2 x}},\qquad
\end{align}
where
\begin{equation}
x=c(s-s_0),\qquad
\label{betaxast}
c=a_2\sqrt{a_1^2-a_3^2},\qquad
\sn x_\ast=\sqrt{1-\frac{a_3^2}{a_1^2}}.
\end{equation}

By using the relation (\ref{TinY}),
the Nahm data corresponding to $Y^a(s)$ of the form (\ref{ellY12})
are obtained as
\eqb\label{Tinfell}
T_\alpha^I\Eqn{=}
  \left((-1)^{\alpha-1}f_Jf_K-f_If_0\right)\frac{\sigma^I}{2}
 +\frac{n_I}{4}\left(f_I^2-f_J^2-f_K^2+f_0^2\right)\bvec{1}_2,
\eqe
where $I,J,K$ are any permutation of $1,2,3$
and $n_I$ are given as in (\ref{nIinthetaphi}).
Substituting (\ref{finwp}) for $f_i$,
the corresponding Nahm data are obtained as
\[
T_\alpha^I=
  \wp_I\left(s-s_\alpha\right)\frac{\sigma^I}{2}
+\frac{n_I}{4}\left(a_I^2-a_J^2-a_K^2\right)
  \bvec{1}_2
\]
with
\[
s_1=s_0-s_\ast,\qquad s_2=s_0+s_\ast
\]
and $a_I$ given as in (\ref{alphaIdef}).
One can also express these Nahm data in terms of
either Jacobi theta functions or Jacobi elliptic functions:
This is done by using
\[\label{wpthetarel}
\wp_I(\tilde{s})
=\frac{\pi\varth_{J+1}\varth_{K+1}}{2\omega_1}
 \frac{\varth_{I+1}(u)}{\varth_1(u)}
\]
with $u=\tilde{s}/2\omega_1$ or
\[\label{wpsncndnrel}
\wp_1(\tilde{s})=c\frac{\cn x}{\sn x},\qquad
\wp_2(\tilde{s})=c\frac{\dn x}{\sn x},\qquad
\wp_3(\tilde{s})=c\frac{1}{\sn x}
\]
with $x=c\tilde{s}=a_2\sqrt{a_1^2-a_3^2}\tilde{s}$, respectively.

Clearly, the above Nahm data are the most general solution
to the Nahm equations with $N=2$.
Up to the $\grp{SU}(2)\times\grp{SO}(3)$
automorphism, either of the two sets of Nahm data
possess six free parameters $x_\alpha,\omega_1,\omega_3,t^I$.
These parameters completely determine the six free parameters
of $Y^a(s)$ with $N=2$.
Thus, we again find that one of the two sets of Nahm data
essentially determine a solution of the BPS equations.

Finally, let us see some particular limits of the solution.
The Weierstrass $\wp$-function is doubly periodic
with respect to fundamental periods $2\omega_1,2\omega_3$.
(Recall that $\tau=\omega_3/\omega_1$
and here we take $\omega_1\in\bbR_{>0}$
and $\omega_3\in i\>\!\bbR_{>0}$.)
One obtains rational solutions by sending either of the periods
to infinity.
If we take the limit $\omega_3\to i\infty$
(i.e.~$\tau\to i\infty$),
we obtain
\[
\label{ratsoltrig}
f_0=\frac{\sin x}{\sin x_\ast}f_2,\quad\
f_1=\frac{\cos x}{\cos x_\ast}f_2,\quad\
f_2=f_3=
  \sqrt{\frac{c\sin x_\ast \cos x_\ast}
             {\sin(x_\ast+x)\sin(x_\ast-x)}},
\]
where
\[
x=c(s-s_0),\qquad
x_\ast=cs_\ast,\qquad
c=\frac{\pi}{2\omega_1}.
\]
On the other hand, if we take the limit $\omega_1\to\infty$
(i.e.~$\tau\to i0$),
we obtain
\[
\label{ratsolhyp}
f_0=\frac{\sinh x}{\sinh x_\ast}f_1,\quad\
f_3=\frac{\cosh x}{\cosh x_\ast}f_1,\quad\ 
f_1=f_2=
  \sqrt{\frac{c\sinh x_\ast \cosh x_\ast}
             {\sinh(x_\ast+x)\sinh(x_\ast-x)}},
\]
where
\[
x=c(s-s_0),\qquad
x_\ast=c s_\ast,\qquad
c=\frac{\pi i}{2\omega_3}.
\]
Note that $\omega_3/i\in\bbR_{>0}$.
Note also that the semi-infinite solution is obtained
from $\tilde{f}_i$ in the limit $\omega_1\to\infty$.

\section{Reduction in connection with periodic Toda chain}

It has been known that the Nahm equations reduce
to Toda molecule equations \cite{Ward:1985ww}.
Here we consider the reduction to the differential
equations for the periodic
Toda chain \cite{Sutcliffe:1996hx} as an example
and present the corresponding reduction
of the BPS equations.
Interestingly, the reduction takes a remarkably simple form
in terms of the BPS equations.

Let us make an ansatz of matrices $Y^a$ as follows:
\[
\label{Todareduction}
(Y^1)_{mn}=g_m(s)\delta_{m,n},\qquad
(Y^2)_{mn}=h_n(s)\delta_{m,n+1},
\]
where $m,n=1,\ldots,N$. In this section the values of indices are
identified mod $N$.
By using a constant $\grp{SU}(N)\times \grp{SU}(N)\times\grp{U}(1)$
rotation one can take all $g_m,h_m$ to be real functions
without loss of generality.
The BPS equations (\ref{BPSeqs}) in this case become
\eqb
\label{gdiffeq}
\dot{g}_m\Eqn{=}\left(h_{m-1}^2-h_m^2\right)g_m,\\
\label{hdiffeq}
\dot{h}_m\Eqn{=}\left(g_{m+1}^2-g_m^2\right)h_m.
\eqe
Next, let us introduce
\begin{align}
\hspace{7em}
a_m&:=g_{m+1}h_m,&
\tilde{a}_m&:=g_m h_m,\nn\\
b_m&:=g_m^2-h_m^2,&
\tilde{b}_m&:=g_m^2-h_{m-1}^2.
\hspace{7em}
\end{align}
These particular combinations
satisfy the differential equations of the periodic Toda chain
with imaginary coupling
\eqb
\dot{a}_m\Eqn{=}a_m(b_{m+1}-b_m),\\
\dot{b}_m\Eqn{=}-2(a_m^2-a_{m-1}^2).
\eqe
The same equations hold for $\tilde{a}_m,\tilde{b}_m$.

The simplest case is the reduction of the form (\ref{Todareduction})
with $N=2$.
In this case, the general solution is given by
the general elliptic solution (\ref{ellY12})
with $\theta=\phi=0$
and $f_i$ given in the last section.

\section{Discussion}

In this paper we have investigated classical integrability of
BPS equations in the ABJM theory.
The integrable structure of the BPS equations is
closely related to that of the Nahm equations.
Making use of this fact, we have formulated an efficient way
of constructing solutions of the BPS equations from those of
the Nahm equations.
As an illustration, we have constructed explicitly
the most general solutions describing two M2-branes suspended
between two parallel M5-branes
as well as two semi-infinite M2-branes ending on an M5-brane.

We have observed that solutions of the BPS equations are
uniquely determined if the corresponding Nahm data are given.
It would be of interest if one could formulate a more direct way of
constructing solutions of the BPS equations from given Nahm data.

We have elucidated that the number of free parameters
of the semi-infinite solutions is in perfect agreement with
the moduli counting of the ABJM theory. On the other hand,
we have not yet found a clear explanation of
the physical meaning of the number of free parameters in the general
solution from the point of view of M-theory.
It is interesting to clarify the structure of
the moduli space of solutions and interpret
it in the context of M-theory.

We have focused on the BPS equations which determine
the ``flat'' M2-branes suspended between parallel M5-branes.
It would be of great interest if BPS equations corresponding to
other configurations could also exhibit a similar integrable structure.

\vspace{3ex}

\begin{center}
  {\bf Acknowledgments}
\end{center}

The authors would like to thank
S.~Kakei, K.~Hosomichi, M.~Hamanaka, S.~Lee, D.~Muranaka,
T.~Nosaka, K.~Takasaki, T.~Takebe for discussions.
The works of K.S.~and S.T.~are supported in part by Grant-in-Aid
for Scientific Research from the Japan Ministry of Education, Culture, 
Sports, Science and Technology (MEXT).

\vspace{3ex}


\appendix

\section{General Nahm data with $N=2$}

In this appendix we present a construction of the general solution
of the Nahm equations (\ref{Nahmeq}) with $N=2$.
The Nahm data $T^I\ (I=1,2,3)$ are hermitian matrices.
For $N=2$, any $T^I(s)$ can be expressed as
\[
T^I(s)=F^I{}_J(s)\frac{\sigma^J}{2}+t^I(s)\bvec{1}_2,
\]
where $F^I{}_J,\, t^I$ are real functions and $\sigma^J$ are
Pauli matrices.
The Nahm equations imply that
\[
\label{Nahmfconst}
\sum_{I=1}^3F^I{}_J F^I{}_K=\mbox{const.},\qquad
\sum_{I=1}^3F^J{}_I F^K{}_I=\mbox{const.}\qquad
\mbox{for}\qquad J\ne K
\]
and
\[
t^I=\mbox{const.}
\]
Recall that the Nahm equations are invariant under
the transformations $T^I\to L^I{}_JT^J$ and
$T^I\to U T^I U^\dagger$,
where $(L^I{}_J)$ and $U$ are constant $\grp{SO}(3)$ and $\grp{SU}(2)$
matrices, respectively.
Using these transformations,
one can transform $F^I{}_J$ at a fixed value of $s$
into the form
\[
F^I{}_J=F^I\delta^I_J.
\]
At this value of $s$,
\[\label{Nahmfoffdiag}
F^I{}_J=0\qquad \mbox{for}\qquad I\ne J.
\]
It then follows from (\ref{Nahmfconst}) that
(\ref{Nahmfoffdiag}) hold for any value of $s$.
Thus, without loss of generality one can assume
the form of the Nahm data to be
\[\label{Nahmgenform}
T^I(s)=F^I(s)\frac{\sigma^I}{2}+t^I\bvec{1}_2,
\]
where $F^I(s)$ are real functions and $t^I$ are real constants.

It follows from the Nahm equations that $F^I(s)$ satisfy
the following differential equations
\[
\dot{F}^I=-F^JF^K,
\]
where $I,J,K$ are any permutation of $1,2,3$.
The general solution to this equation is most concisely expressed
in terms of Weierstrass elliptic functions as
\[
F^I=\wp_I(s-s_0;2\omega_1,2\omega_3)
\]
with
\[
s_0\in\bbR,\qquad\omega_1\in\bbR_{>0},\qquad\omega_3\in i\bbR_{>0}.
\]
It is also possible to express $F^I$ in terms of
Jacobi theta functions or Jacobi elliptic functions
by using (\ref{wpthetarel}) or (\ref{wpsncndnrel}),
respectively.
Any permutation of $F^I$ and/or
overall sign change of 
even number of $F^I$ again give another solution
to the equations.

\section{Conventions of special functions}

The Jacobi theta functions are defined as
\eqb
\varth_1(z,\tau)\Eqn{=}
 i\sum_{n\in \bbZ}(-1)^n y^{n-1/2}q^{(n-1/2)^2/2},\\
\varth_2(z,\tau)\Eqn{=}
 \sum_{n\in \bbZ}y^{n-1/2}q^{(n-1/2)^2/2},\\
\varth_3(z,\tau)\Eqn{=}
 \sum_{n\in \bbZ}y^n q^{n^2/2},\\
\varth_4(z,\tau)\Eqn{=}
 \sum_{n\in \bbZ}(-1)^n y^n q^{n^2/2},
\eqe
where $y=e^{2\pi i z},\ q=e^{2\pi i \tau}$.
We often use the following abbreviated notation
\[
\varth_k(z):=\varth_k(z,\tau),\qquad \varth_k := \varth_k(0,\tau).
\]
The Weierstrass $\wp$-function is defined as
\[
\wp(z)=\wp(z;2\omega_1,2\omega_3)
:=\frac{1}{z^2}
+\sum_{(m,n)\in\bbZ^2_{\ne (0,0)}}
\left[\frac{1}{(z-\Omega_{m,n})^2}
  -\frac{1}{{\Omega_{m,n}}^2}\right],
\]
where $\Omega_{m,n}=2m\omega_1 + 2n\omega_3$.
We also introduce the following notation
\[
e_I:=\wp(\omega_I)\qquad (I=1,2,3),
\]
with
\[
\omega_1+\omega_2+\omega_3=0,\qquad
\frac{\omega_3}{\omega_1}=\tau.
\]
The functions
$\wp_I(z)=\wp_I(z;2\omega_1,2\omega_3)$ are then defined as
\[
\wp_I(z):=\left(\wp(z)-e_I\right)^{1/2},
\]
where the branch of the square root is chosen
so that $\wp_I(z)=1/z+{\cal O}(z)$.
Let us present some useful formulas:
\eqb
\wp'(z)\Eqn{=}-2\wp_1(z)\wp_2(z)\wp_3(z),\\[1ex]
\wp_I(z+w)\Eqn{=}
-\frac{\wp_I(z)\wp_J(w)\wp_K(w)-\wp_I(w)\wp_J(z)\wp_K(z)}
      {\wp(z)-\wp(w)},
\eqe
where $I,J,K$ are any permutation of $1,2,3$.
The following relations are also useful:
\[
\wp_1(s)=\sqrt{e_1-e_3}\frac{\cn x}{\sn x},\qquad
\wp_2(s)=\sqrt{e_1-e_3}\frac{\dn x}{\sn x},\qquad
\wp_3(s)=\sqrt{e_1-e_3}\frac{1}{\sn x},
\vphantom{\bigg|}
\]
\[
\wp_I(s)
=\frac{\pi}{2\omega_1}\frac{\varth_2\varth_3\varth_4}{\varth_{I+1}}
 \frac{\varth_{I+1}(u)}{\varth_1(u)},
\]
where
\[
x
=\pi\varth_3^2 u
=\sqrt{e_1-e_3} s.
\]

\vspace{3ex}


\renewcommand{\section}{\subsection}
\renewcommand{\refname}

\end{document}